\documentclass[conference]{IEEEtran}
\usepackage{cite}
\usepackage [para]{threeparttable}
\usepackage[hyphens]{url}
 \usepackage{array,multirow,graphicx}
\usepackage{float}
\usepackage{makecell}

\ifCLASSINFOpdf
\else
\fi
\begin{document}
	\sloppy
	\pagestyle{plain}
%
\title{Empirical Analysis of Factors and their Effect on Test Flakiness - Practitioners' Perceptions}

\author{\IEEEauthorblockN{Azeem Ahmad, Ola Leifler, Kristian Sandahl}
\IEEEauthorblockA{Department of Computer Science\\
Linköping University, Sweden\\
Email: firstname.lastname@liu.se}
}


%


\maketitle

\begin{abstract}
Developers always wish to ensure that their latest changes to the code base do not break existing functionality. If test cases fail, they expect these failures to be connected to the submitted changes. Unfortunately, a flaky test can be the reason for a test failure. Developers spend time to relate possible test failures to the submitted changes only to find out that the cause for these failures is test flakiness. The dilemma of an identification of the real failures or flaky test failures affects developers' perceptions about what is test flakiness. Prior research on test flakiness has been limited to test smells and tools to detect test flakiness. In this paper, we have conducted a multiple case study with four different industries in Scandinavia to understand practitioners' perceptions about test flakiness and how this varies between industries. We observed that there are little differences in how the practitioners perceive test flakiness. We identified 23 factors that are perceived to affect test flakiness. These perceived factors are categorized as 1) Software test quality, 2) Software Quality, 3) Actual Flaky test and 4) Company-specific factors. We have studied the nature of effects such as whether factors increase, decrease or affect the ability to detect test flakiness. We validated our findings with different participants of the 4 companies to avoid biases. The average agreement rate of the identified factors and their effects are 86\% and 86\% respectively, among participants.
\end{abstract}


%
\IEEEpeerreviewmaketitle

\section{Introduction}
Regression testing, automatic or manual, makes sure that a change made in one part of the system does not break another part of the system (e.g., break existing functionality). Developers submit code changes and expect possible test failures to be connected with the submitted change. Unfortunately, some test failures are not due to the submitted changes but flaky tests. In addition to this, tests failing without any change in the code base (e.g., regression tests executing on the same build) are also called flaky tests. The most common definition of a flaky test, in literature, is: \textit{a test that exhibits both pass and failure outcomes when no changes are introduced in the code base}. King et al. extend this definition\cite{king_towards_2018}: \textit{"flaky tests exhibit both passing and failing results when neither the code nor test has changed"}. Flaky tests are also called\textit{ "unreliable tests whose outcome is not deterministic"} \cite{palomba_does_2017}. The latter definition does not provide any context whether the test case's unreliability is associated, in any way, with the system under test (SUT), test case code or environment changes.  

As discussed above, we have noticed differences in definitions of test flakiness (TF). This is one of the reasons that larger software continues to suffer from a high amount of test flakiness. For example, flaky tests at Google are account for 4.56\% of the total 1.6M failed test cases \cite{luo_empirical_2014}. It has also been reported that 13\% \cite{labuschagne_measuring_2017} and half \cite{hilton_trade-offs_2017} of all builds failed due to test suites that contains flaky tests. one in seven test cases written by engineers, at Google, fail due to test case flakiness \cite{noauthor_flaky_nodate}. There is also active discussions about TF on several blogs \cite{noauthor_flaky_nodate,gang_flaky_2017,noauthor_eradicating_nodate,noauthor_flakiness_nodate,noauthor_no_2012}. 

The research conducted in TF is either empirical analysis (e.g., studies on open source software) of test smells \cite{palomba_does_2017,luo_empirical_2014,thorve_empirical_2018}, tools or techniques to detect test smells related to TF \cite{bell_deflaker:_2018,shi_detecting_2016,gyori_nondex:_2016,gambi_practical_2018}, machine learning approaches to predict TF \cite{king_towards_2018} or attracting researcher's attention towards TF \cite{rahman_impact_2018}. Given the differences, as discussed above, in defining TF, We have seen no study that captures practitioner' perceptions of TF. This is the first study that identifies factors (other than test smells) that affect TF. In addition to this, all related work mentioned in Section \ref{relatedwork} is conducted with open-source software whereas we are conducting this research in close collaboration with four Scandinavian industries. 

The need to understand practitioners' perceptions with respect to software engineering has received great attention from the scientific community. Researchers have investigated practitioners' perceptions of continuous integration\cite{laukkanen_stakeholder_2015}, software programming (e.g., code smells or exception handling)\cite{sun_effectiveness_2016,ebert_study_2013,shah_understanding_2010,palomba_they_2014}, software testing\cite{camacho_agile_2016,percival_developer_2013,tan_test_2018} and software quality\cite{bavota_empirical_2013,wan_perceptions_2018,zou_how_2018,abad_task_2017} and observed differences in practitioners' perceptions of subject under investigation. We have found no study that understand practitioners' perceptions with respect to TF. Practitioners, sometimes, favor local opinion over empirical evidence of new techniques' adoption which makes perception important \cite{rainer_persuading_2003}.

We present a multiple case study, with four different companies, with the objective of understanding practitioners' perceptions about TF such as how they define test flakiness as well as what factors (other than code smells), in practitioner's perception, can increase, decrease or affect the ability to detect TF. We validated our findings during workshop and map factors to available literature (e.g., factor's categorization). 

Section \ref{background} presents the background, problem domain of this study and common definitions/concepts. Related work is presented in Section \ref{relatedwork}. The Research protocol is presented in Section \ref{researchm}. Section \ref{result} presents qualitative and quantitative results and practitioners' percetions about TF is discussed in Section \ref{discussion}. Validity threats and conclusion can be found in Section \ref{validtythreasts} and \ref{conclusion} respectively.

\section{Background and Problem Domain}
\label{background}
Flaky tests are very common in large code bases. The current approaches to handling TF are not satisfactory \cite{luo_empirical_2014}. Test flakiness prediction, assessment, and reduction are identified as open problems that have yet to receive significant attention from the scientific community \cite{harman_start-ups_2018}. We have observed that current studies mostly depend on software artifacts (e.g., source code and test cases) which is just a small part of the bigger problem and core issues are still not addressed for TF. Authors, in this study, take a step back to investigate the perceptions of TF among practitioners and identify factors that increase, decrease or affect the ability to detect TF. This is a unique study focusing on understanding practitioner's perceptions and what factors affect test case flakiness. We categorized the identified factors into four distinct categories. It is worth mentioning here that ''re-run", as mentioned in \cite{bell_deflaker:_2018} is the most widely used technique to address test flakiness.

According to an online dictionary, "the perception is the way, in which something is regarded, understood or interpreted" i.e., a mental impression. Since the participants of this study are experienced is software technology, their perception, about TF, is likely to be influenced by what they have heard, observed and experienced in the workplace. Perceptions may not be a true assessment of reality but we consider it very important. 

The difference between factors and code smells can also be described through a literal definition. Factor, as defined in an online dictionary, is a circumstance, fact, or influence that contributes to a result. The code smell can be one factor.  

smells refer to any characteristic in the programming code that possibly indicate a problem. Code smells refer to smells in source code or system under test whereas test smells refer to smells in the test case code.

We have observed that flakiness can be found in SUT source code or some practitioners believe that flakiness can be found in external dependencies so whenever we refer to test flakiness (TF), we cover every aspect of flakiness although the name can be confused with the "\textbf{test}" word flakiness. We keep switching between the TF or test case flakiness to enhance the reader's readability but both carry the same meaning.

\section{Related Work}
\label{relatedwork}

Luo et al. in \cite{luo_empirical_2014} categorized the causes of test case flakiness by investigating 52 open-source projects and 201 commits. Asynchronous wait (45\%), concurrency(20\%) and test order dependency (12\%) were found to be the most common causes of TF. The suggestions, provided by Luo et al. are to avoid specific code smells that lead to TF. Results, presented by Luo et al. are partially replicated by Palomba and Zaidman in \cite{palomba_does_2017} leading to the conclusion that most prominent causes of TF are asynchronous wait, concurrency, and input-output issues. Palomba and Zaidman report that network issues (10\%) are also one of the reasons for TF. Another empirical study of root causes of TF in Android Apps was conducted by Thorve et al. \cite{thorve_empirical_2018} by analyzing the commits of 51 Apache open-source projects. Thorve et al. \cite{thorve_empirical_2018} complement the results of Luo et al. and Palomba and Zaidman but also reported two additional test smells (e.g., user interface and program logic), related to TF in Android Apps.

Bell et al. in \cite{bell_deflaker:_2018} proposed a new technique called DeFlaker which monitors the latest code coverage and marks the test case as flaky if the test case does not execute any of the changes. Another technique called PRADET \cite{gambi_practical_2018} does not detect flaky test directly but uses a systematic process to detect problematic test order dependencies. These test order dependencies can lead to TF.  

King et al in \cite{king_towards_2018} present an approach that leverages Bayesian network for flaky test classification and prediction. This approach considers flakiness as a decease that can be mitigated by analyzing the symptoms and possible causes. Teams, using this technique, improved CI pipe-line stability by as much as 60\%. 

We did not find any study, to best of our knowledge, that investigates factors and their effect on test flakiness.

\section{Research Methodology}
\label{researchm}
\begin{figure}
	\centering
	\includegraphics[width=\columnwidth]{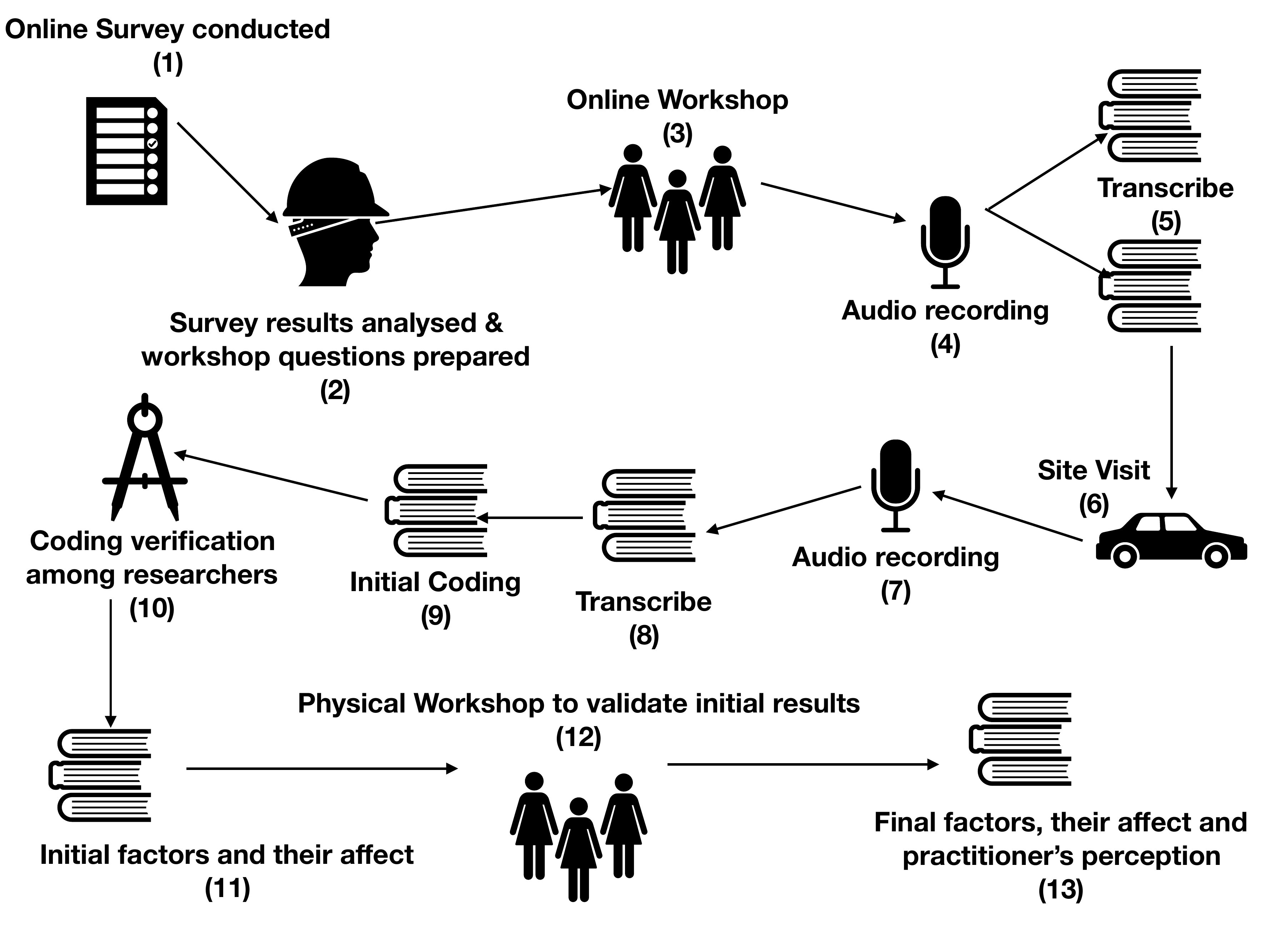}
	\caption{Research Methodology Protocol for Data Collection, Analysis and Validation.}
	\label{fig:methodology}
\end{figure}

We presented a multiple-case holistic study \cite{yin_case_2009}. We studied one phenomenon (e.g., test flakiness) in four companies (e.g., cases). Each case, in multiple case study, should be selected carefully so that it either a) predicts similar results or b) predicts contrasting results \cite{yin_case_2009}. We picked (a) so that results from different cases should complement to each other in order to enhance our understanding about practitioner's perception about TF rather than contradicting with each other. Yin in \cite{yin_case_2009} concluded that multiple case study provides more compelling and robust evidence that can be generalized to a greater extent than the findings from a single case study. Figure \ref{fig:methodology} presents the research methodology protocol in detail. The numbers (1-13), in Figure 1, represents the sequence of conducted activities to collect, analyze and validate data.
\subsection{Case Description}
\label{studydescription}
All cases as described below involve industries from Sweden and Denmark. We maintain little anonymity when it comes to describing the case companies due to the non-disclosure agreements between companies and our University. 
\subsubsection{Case 1}
Swedish company that involves surveillance equipment. The head office is in Sweden while running businesses all around the world. The investigation focused on people involved in software testing. This case company has a mature automated continuous integration pipeline and most of the testing is done through automation.
\subsubsection{Case 2}
Swedish company that involves automotive equipment. The head office is in Sweden while running businesses all around the world. The investigation focused on people involved in software testing and development. This case company has achieved a good level of testing automation.
\subsubsection{Case 3}
Swedish company that involves medical equipment. The head office is in Sweden while running businesses mostly domestically.  The investigation focused on people involved in software testing and development. This case company is shifting towards test automation.
\subsubsection{Case 4}
Danish company that involves water equipment. The head office is in Denmark while running businesses all around the world. The investigation focused on people involved in software testing (e.g., embedded systems), management and development. This case company has achieved a good level of automation.
\subsection{Research Question}
\label{researchquestions}
We address the following research question in this study to understand practitioner's perception about TF and what and how different factors affect TF:\\

\textit{RQ1: What factors do practitioners perceive affect the test flakiness?}

There are three important things in RQ1. First, we need to investigate the factors that affect test flakiness. Second, we need to find out the magnitude of the effect, for example, investigate if factors increase, decrease or affect the ability to detect test flakiness. Third and last, we need to understand the practitioners' perceptions of identified factors and their effect.

\subsection{Data Collection}
\label{datacollection}
We collected data, in all four cases, through group interviews, workshop and semi-structured interview. Most of the interactions were conducted online through Zoom or Skype. We recorded the conversation and took notes during the workshop with prior permission from the participants. Table I presents detailed information about data collection.

First, we conducted an initial survey to investigate if case companies are interested in TF through a google form (1 in Figure 1). The Survey\footnote{\sloppy\url{ https://docs.google.com/forms/d/e/1FAIpQLSeOk8YswlfdeEmmRJsVZRCYI_0zwQiRb1RGOinXKVLpkyovew/viewform} } can be accessed online. The questions of survey were prepared by analyzing the prior literature work in test flakiness to understand the 1) common causes of test flakiness in industry, 2) how practitioners write test cases, 3) if there are testing review strategies that includes test flakiness and 4) how much time, it takes to identify and resolve test flakiness. The survey was answered by 10 participants of 5 different companies but only 4 companies proceeded further with this study. After the online survey, we analyzed the results and prepared questions for the workshop (2 in Figure 1). We conducted 120 minutes long workshop to collect data from case 1, 2 and 4 (3 in Figure 1). This workshop was conducted online through Zoom and seven participants (5 testers and 3 developers) from three companies participated. We recorded the audio and later transcribed into an Excel sheet (4 and 5 in Figure 1). Similar online workshop of 120 minutes was conducted with case company 3 on different occasion. The set of questions\footnote{\sloppy\url{https://docs.google.com/document/d/1cPIFhd9B3DWUXzVbzJ0lJW8nlcIoN0VOjmlHPzkXXME/edit?usp=sharing}} to be discussed during the workshop, can be accessed online.  
Later, one of the authors visited case company 1 to conduct an interview with 1 tester and 1 developer (6 in Figure 1). The interview was 180 minutes long and recorded and transcribed into an excel sheet for analysis (7and 8 in Figure 1). The motivation to visit only one case company was to look at their testing process and documentation, in person, due to their claim of having very low test flakiness. The idea was to extract some hidden knowledge that are embedded in the day to day practices of the company.
\begin{table}[ht]
	\begin{threeparttable}
		\caption{Data Collection Protocol}
		\centering 
		\begin{tabular}{p{0.5cm} p{2.8cm} p{0.4cm} c p{2cm}}
			\hline\hline                        
			Case & Type & Length (Min) & Date & Participants Designation \\ [0.5ex]
			\hline        
			1,2,3,4 & Survey & - & 2018-10  & 6\tnote{T}, 4\tnote{D} \\
			1,2,4 & Workshop & 120 & 2018-11  & 5\tnote{T}, 3\tnote{D} \\
			3 & Workshop & 120 & 2018-12  & 1\tnote{T}, 1\tnote{D} \\
			1 & Interview, Site Visit & 180 & 2018-12  & 2\tnote{T} \\
			
			\hline
		\end{tabular}
		\begin{tablenotes}
			\footnotesize
			\item[T] Test Lead \item[D] Developer
		\end{tablenotes}
	\end{threeparttable}
\end{table}
\subsection{Preparation}
\label{preparation}
During data collection, two authors took notes in addition to an audio recording. We anonymized the data in such a way that it should not be traced back to the individual participants. We obtained the informed consent from participants for audio recording. We transcribed the audio recording word-by-word into an excel sheet using an audio player. Audio clips were played 3 times, before writing the text into excel sheet. Each cell in Excel sheet contains text from one person during conversation, labeled with who said what. The names were also anonymized with PX, where X is the number assigned by us to participant for our understanding. Two of the authors check the transcription to find any discrepancies or missing information. 
\subsection{Data Analysis}
\label{dataanalysis}
We read all the transcript from workshop and interview and coded according to open coding \cite{strauss_basics_1998}. Each paragraph of the transcript of each case company was studied to determine what was said and each paragraph was labeled with one or more codes. Later, we compared all paragraphs from different case companies to collect similar codes (axial codes). These codes were further divided as the influencing factor and their effect such as whether the factors increase, decrease or affect the ability to detect TF. Table II presents two examples of how we extracted influencing factors and their effect on TF. We have used three different symbols to denote the effect of influencing factors in TF. Plus sign (+) represents an increase, minus sign (-) represents a decrease and star sign (*) represents a detection.

\begin{table*}[ht]
	\caption{Two examples to extract influencing factors and their effect on TF from text}
	\centering
	\begin{tabular}{p{0.04\linewidth}p{0.27\linewidth}p{0.24\linewidth}p{0.14\linewidth}p{0.19\linewidth}}
		\hline
		\hline                     
		- & Text & Identified Factor & Effect & Textual Description \\
		\hline   
		Ex: 1 & “The reason for flakiness for the project we have now is, where we lack requirements and the reason was test case code. Some of the changes were made in the test case. This is for undefined project and unclear requirements” & Lack of clear requirements & + TF (increase test flakiness) &  Lack of clear requirements increase test flakiness \\ \hline
		Ex: 2 & “network is very interesting and one thing that is on my to-do list it that we are trying to do in a very wrong way. We are trying to avoid network issues by making sure that the network is up and running. If you assume that network /ethernet network/IP network is perfect you will have problems and the way we are approaching is to make it perfect have switches and making sure that XXXX hardware is ok and making sure that everything is correct. What Netflix is doing by actively undermining the network, by making it worse, you design everything to handle the problem instead of trying to clean the ways for them. We are cleaning away the problem not by designing to handle it well/back.” &  \makecell[l]{Robust test case  \\ Undermining network infrastructure}
				&    - TF (reduce test flakiness) & Test flakiness is reduced when you develop robust test case that handles uncertainties and you intentionally undermine the network performance to check the test case robustness \\
	
		\hline
	\end{tabular}
\end{table*}

\subsection{Data Validation}
\label{datavalidation}
After analysis, the initial codes were checked by two of the researchers. Later, we conducted a physical workshop to validate results with seven different participants from three industries. We intentionally selected different participants for data validation than those who participated earlier (e.g., data collection) in order to avoid generalization and biases in results. This workshop was conducted for 120 minutes and participants were provided with the list of influencing factors together with their descriptions. Participants were asked to rank identified factors based on High, Low and Do not Agree scale. Each factor was discussed among participants to understand it. In addition to factor ranking, participants were asked to rank the effect of influencing factors on test flakiness. Participants were provided with a Likert scale (e.g., 1 to 5 - Strongly Agree to Strongly Disagree) to give their opinion.
\section{Results}
\label{result}
We identified 23 unique factors. The example of how we extracted these factors can be found in Section \ref{dataanalysis}. This section explains each factor in detail together with the practitioner's quote and assigned category. We do not intend to provide all the quotes, except when needed. We also explain how a factor affect TF in Table IV. Table III presents the category, the identified factors and their mapping to the available literature. We did not find any available literature for the company-specific factors (e.g., last row). One can easily determine, by looking at Table III, that all identified factors have been mentioned in the literature but in different contexts such as what is a good test case or how to improve software quality. The categorizations were achieved in the context, in which the factors were discussed. For example, simple test case (i.e., identified factor in this study) is mentioned as 'simplicity' in \cite{bowes_how_2017} in the context of what are good tests. We will discuss Table IV contents more in detail in Section \ref{discussion}.
\begin{table*}[ht]
	\label{categogryAndFactor}
	\caption{Identified Factors, their mapping to available literature and category	}
	\centering
	\begin{tabular}{p{0.15\linewidth}p{0.27\linewidth}p{0.29\linewidth}}
		\hline
		\hline                     
		Category & Mapping of Identified Factor to Available Literature & Identified Factor \\
		\hline   
		{\multirow{9}{*}{\rotatebox[origin=c]{0}{Software Test Quality}}} & Simplicity \cite{bowes_how_2017} & Simple test case  \\
		& Single responsibility
		\cite{bowes_how_2017} 		 & Small test case \\
		& Single responsibility \cite{bowes_how_2017} 			 & Test case testing in a specific way \\
		& Obsolete test case \cite{bowes_how_2017} 		 & Older test case \\ 
		& Fast feedback	\cite{bowes_how_2017} 	 & Longer running system under test or test case \\ 
		& Simplicity of fixtures \cite{bowes_how_2017}\cite{rompaey_characterizing_2006}	& Handler outside test cases	  \\ 
		& Asses conformance to regulation\cite{kaner_what_nodate}\cite{beer_initial_2017}			 & Lack of clear requirements \\ 
		& Robust test case \cite{deursen_refactoring_2001}				 & Robust test case	\\ 
		& Configuration Issues \cite{vahabzadeh_empirical_2015}				 & Lack of environment understanding\\
		& Complex System \cite{bavota_are_2015}  & Complex System\\
		
		\hline
		{\multirow{3}{*}{\rotatebox[origin=c]{00}{Software Quality}}}& Robust network \cite{deursen_refactoring_2001}	 &  Undermining network infrastructure \\
		& Assure Quality \cite{kaner_what_nodate}		 & Team motivation		  \\
		& Experienced team \cite{kaner_what_nodate}				 & Experienced team	 \\
		\hline
		{\multirow{4}{*}{\rotatebox[origin=c]{00}{Known Flaky Test}}}& Test (in)dependency \cite{bowes_how_2017} 	 &  Higher dependency among test case \\
		& Test code smells	\cite{deursen_refactoring_2001}				 & Test code smells	  \\
		& Better test results reporting	\cite{deursen_refactoring_2001}					 & Better test results reporting	 \\
		& Rerun test cases	\cite{bell_deflaker:_2018}					 & Rerun test cases \\
		\hline
		{\multirow{6}{*}{\rotatebox[origin=c]{00}{Company Specific}}}& - &  Test the test \\
		&  - 	 & Testing for flaky test on different levels \\
		& - 						 & Hard coding	  \\
		& - 			 & Unstable CI release \\
		& - 							 & Web component run-time generation  \\
		& - 							 & Complex feature avoidance  \\
		\hline
	\end{tabular}
\end{table*}
\begin{table*}[ht]
	\caption{Wide single-column table in a twocolumn document.}
	\centering
	\begin{tabular}{p{0.25\linewidth}p{0.07\linewidth}p{0.60\linewidth}}
		\hline
		\hline                     
		Identified Factor & Effect on test flakiness & Textual description of combined/unique factors and affect\\
		\hline   
		Test the test & * TF & Test flakiness is detected when you test the test. e.g., A test Q that tests action Q is a flaky test.  Then you can write a test that explicitly catches this flakiness, e.g. a test that repeats action Q 100 times and explicitly states that it needs to succeed 100 times
		\\
		\hline
		Complex system & + TF & Test flakiness is increased when the system is very complex AND you do not understand the environment (e.g, how the system under test interact with network, ports or external libraries)
		\\
		\hline
		Lack of environment understanding & + TF & Test flakiness is increased when you do not understand the environment (e.g, how the system under test interact with network, ports or external libraries)
		\\
		\hline
		Testing on different level & - TF & Test flakiness is reduced when there are many lines of defenses before releasing the product/firmware/API specifically targeting flaky tests.
		\\
		\hline
		Re-running test cases & * TF & Test flakiness can be detected by re-running test cases many times on all product on different occasions (overnight or weekend)
		\\
		\hline
		Hard coding & + TF & Test flakiness is increased when you hard coded if-else conditions to handle specific situations
		\\
		\hline
		Test smells & + TF & Test flakiness is increased when there are code smells (e.g., thread.sleep, wait, race conditions) 
		\\
		\hline
		Experienced team & - TF & Test flakiness is increased when a team does not have previous experience in handling test flakiness
		\\
		\hline
		Unstable CI release & + TF & Test flakiness is increased with unstable Jenkins release and no ability to rollback to previous Jenkins version. In this case, you just have to wait until Jenkins updates
		\\
		\hline
		Robust test case & - TF & Test flakiness is reduced when you develop a robust test case that handles uncertainties 
		\\
		\hline
		Undermining the network infrastructure & - TF & Test flakiness is reduced you intentionally undermine the network to check the test case robustness
		\\
		\hline
		Environment handlers outside test cases & - TF & Test flakiness is reduced when complex codes are removed from test cases and placed in external libraries
		\\
		\hline
		Small test case & - TF & Test flakiness is reduced when test cases are small (e.g., lines of code) 
		\\
		\hline
		Simple test case & -TF & Test flakiness is reduced when test cases are simple (1 -2 assertions) 
		\\
		\hline
		Test case testing in a specific way & - TF & Test flakiness is reduced when you test the functionality is specific ways (e.g., testing different product behaviors or states should be done with different test cases) 
		\\
		\hline
		Higher dependencies among test cases & + TF & Test flakiness is increased when there is a higher dependency among test cases and these dependencies are not explicitly documented
		\\
		\hline
		Better reporting test results & * TF & Test flakiness can be detected with better test results reports or test result log files or comparing test execution histories
		\\
		\hline
		Older test cases & + TF & Test flakiness is increased when product functionality has been updated and test cases are still old / not updated
		\\
		\hline
		Custom web components run-time generation & + TF & Test flakiness is increased with custom web components (e.g., drop-down menu or text-views) that are generated run-time
		\\
		\hline
		Team motivation & - TF & Test flakiness is reduced when the team is motivated to hunt it
		\\
		\hline
		SUT /test case longer execution & * TF & Test flakiness is detected when SUT runs for a longer time or test case runs over a longer period of time
		\\
		\hline
		Lack of clear requirements & + TF & Test flakiness is increased with a lack of clear requirements or with undefined project
		\\
		\hline
		Avoiding complex feature/requirements testing & - TF & Test flakiness is reduced when complex functionality is not tested in detail (e.g., sending a restart signal to the product but not making sure if it actually restarted)
		\\
		\hline
	\end{tabular}
\end{table*}
\subsection{Qualitative Results}
\subsubsection{Software Test Quality Related Factors}\hfill\\
\label{softwaretestquality}
All the factors, in this section, are categorized as software test quality due to the fact that these factors have been mentioned in literature in the context of test quality. 

\textbf{Older test case}: Older test cases refer to those test cases that have not been updated, even the corresponding functionalities have been changed, long ago as stated by one of the participants: \textit{"most of the flakiness is because of the system configuration or missing test data or license. Those are often easy to find [....] but when it seems to be the product issue, for example, the product has been changed or you have not updated the test case, that is also more difficult for me as I do not know all the products in detail as the responsible team"}.\\
\textbf{Longer running SUT or test case}: Some functionalities, of product, require longer execution of the software so that the flaky behavior can be understood and detected as one of the participants explained: \textit{"In some cases, even though the software was not changed, the test [flaky] was still running in the night . You could call it a kind of long term testing or long term reliability testing and for this reason, flaky behavior was detected, which was not possible without it [longer execution]"}\\
\textbf{Environment Handler outside test cases}: Complex codes are removed from test cases and placed in external libraries, to reduce test flakiness, as mentioned by one of the participant \textit{"we remove complex code from test case and place it into separate library such as Do-X (actual feature is anonymized due to nondisclosure agreement), all products can use this function differently so we have a library that provides different Do-X function. The test code only says "Do-X" and depending on the product, we can call the function from library"}. The individual test case does not need to know if the network or product is not available and should only fails (e.g., when it is supposed to fail) rather than showing non-deterministic behavior. These external libraries will provide proper test logs result such as 'test failed due to no network availability for product X''.  \\
\textbf{Lack of clear requirements}: Lack of clear requirements and undefined project are mentioned as the main cause of the increase in test flakiness in one of the case company. One of the participants said that \textit{"for the project we have now, where we lack requirements the reason for test flakiness was test case code. Some of the changes were made in a test case. This is for an undefined project and unclear requirement}. Another participant mentioned \textit{"we were not sure [due to lack of requirements] which feature should go in what hardware and that leads to flakiness because people had a different definition at a different stage of the pipeline such as which test case to run [....]"}\\
\textbf{Robust test case}: Robust test cases should be written to address any uncertainty in/around SUT, integration with external libraries or configurations (e.g., network or port issues) thus leading to reduce test flakiness. One of the participants mentioned \textit{whenever we see network problem, we never disable test cases. We do not do anything that is why we see a lot of test flakiness. But later, we started building robust test cases to handle a bad network and we might be as deterministic as we can"}. A test case can become complex if we want to address all uncertainties as shared by one of the participants \textit{"and I will say that it should not be test case [that contain complexity] but somewhere else [handler outside test case]"}.\\
\textbf{Lack of environment understanding}: The word 'environment' refers to SUT, system implementation details, configurations and integration with third-party libraries. The system can be so complex as mentioned by one of the participants \textit{"it is rather that they are not seeing things that are changing in the environment and sometimes the source of this change is a timing issue or network where you can have completely deterministic behavior. For example, a new internee writing a test case without knowing the complete environment and product behavior"}.\\
\textbf{Simple test case}: One of the case company, in this study, writes simple test cases. For example, 1- 2 assertions per test case. This case company claimed to have very low test flakiness. One of the participants said: \textit{"The complexity is interesting. What we have, is a simple test and we try to make it simple and concrete and this is probably what is related to TF. It is kind of a trade-off"}. Another participant explained: \textit{"What is the definition of simple is of-course, not obvious but mostly, you can say it has one assertion and on average less than two"} \\
\textbf{Small test case}: One of the design principle of a single test, in one case company, that it should be small for example, a test case with fewer number of lines. One of the participant shared previous practices: \textit{"It is important because we have before started doing this, all tests were written by people does not have any design principle and they have all type of crazy code. You can have test who has thousands of line of code and some of the code is interacting with X system, which it is not supposed to do due to policies}.\\
\textbf{Test case testing in a specific way}: Each test should test one requirement in a specific way and if the requirements have many facets, many test cases should be written as mentioned by one of the participants \textit{"This small test will know how to test this feature in a specific way and all test should know this"}.
\subsubsection{Software Quality Related Factors}\hfill
 \\ These factors are related to software quality category due to the fact that these factors have been mentioned in literature in the context of software quality. .\\
\textbf{Complex system}: Systems, where things are changing in background without being caught by the naked eye, affect test case flakiness as one of the participants said: \textit{"so the observer is looking at some test that is giving you some information and they are surprised and they are not expecting non-deterministic behavior because they do not see anything changes but system can be so complex that things are changing without them knowing [leading to flakiness]"}. \\
\textbf{Undermining network infrastructure}: We should assume worst case scenarios, when writing test cases for networks, ports, I/O or third-party libraries as mentioned by one of the participants \textit{"We are trying to avoid network issues by making sure that network is always up and running. If you assume that network (Ethernet/ IP network) is perfect, you will have problems [flaky test]"}. Another participant explained this further \textit{"What Netflix is doing - actively undermining the network, by making it worse, so, you design everything to handle the problem instead of trying to clean the ways for them. We are cleaning away the problem not by designing to handle it well"}. \\
\textbf{Team motivation}: Teams should be motivated enough to hunt for TF. One of the participants mentioned \textit{"it was a very persistent analysis when people said that we really want to know why this test case is failing and we spent days to know [it was later found out that this was flaky test case]"}. Another participant shared the same opinion \textit{"one or four times we change the criteria [test] but test case is still flaky then we also debug what could be the problem and at the end, we figured it out [it was flaky]. It was hard to find"}. \\
\textbf{Experienced team}: Previous experiences in test flakiness help as mentioned by one of the participant \textit{"generally I look at test log for how many tests fail then I can generally work out for what causing it then[...] I can tell easily, due to past experiences [if it is flaky] but if I can't then I go to the person who wrote the test and ask what caused it [....]"}. Another participant stated that \textit{"experienced tester that write thread.sleep(60) and they do not know how and why or stuff like that"}. Another participant shared how experience helps in dealing with test flakiness \textit{"[.....] Depending on the how experienced the team is writing the test I say the most of them think about Asynchronous call and wait"}. \\
\subsubsection{Known Flaky Tests Factors}\hfill
\\ These factors are related to flaky tests category. These factors are known to be connected with test flakiness.\\
\textbf{Test smells}: Many different test smells (e.g., thread.sleep(), connection.close() etc) are main reasons for test case flakiness. Our focus is not to list them here but to label it as a factor that increases, decreases or affect the ability to detect TF. One of the participants said:  \textit{"I would say that test code, on average, is as crappy as source code but people pay attention to the source code but for not for test code"} \\
\textbf{Higher dependency among test cases}: This factor complements to what other researchers have concluded that dependency among test cases increase TF as supported by the practices of the case company that has a lower TF \textit{"you can take away test cases so you have very less test case dependency and they are not allowed to depend on another test"}. \\
\textbf{Rerun test cases}: All the participants stated that re-running test cases over a night or weekend helps them detecting TF \textit{"when we write a test case we test it regularly  [..] After we have written it and integrated it so we run it on all products (approx. 50) each night for 7 days two times. It is 2 x 7 x 50 = 700 It is roughly 100 times per night as an order of magnitude"}.  \\
\textbf{Better test results reporting}: A better mechanism to report, log and display test case results help in detecting test case flakiness as mentioned by one of the participants \textit{"as soon as team city fails the test, we have monitor that says which project is failing and you see within which test is failing . We also take screenshots and take logs  [...] Initially it is me that discover that test is failing so I take a look whether it seems based on system itself, system under test or something else [flaky].. if it looks like something else [flaky] then I assign this failure to someone in the team responsible for the test so they can look at it and fix and when it is ok, submit back"}. \\
\subsubsection{Company Specific Factors}\hfill
\label{companyspecific}
\\ These factors are related to the company-specific category. We observed that the following factors may not be applicable to all industries and domains due to fact that these factors are embedded in day to day routines/practices of the specific companies. This is the reason, we did not map these factors to available literature as presented in Table III. \\
\textbf{Testing for flaky test on different levels}: Different testing activities (e.g., unit, system or integration testing) should allow practitioners to detect test case flakiness in the pipeline. The chances to detect test case flakiness increases when you hunt for it as one of the participants said \textit{"The automated tests are not the only line of defense but we have several instances where we expect the flaky test to be caught on the way"}\\
\textbf{Test the test}: if the test exhibits flakiness, you should write another test case that explicitly catches this flakiness, for example, one of the participants mentioned: \textit{"a test Q that tests action Q is a flaky test.  Then you can write a test that explicitly catches this flakiness: a test that repeats action Q 100 times and explicitly states that it needs to succeed 100 times"}. Testing the test will help to detect test case flakiness since 'rerun' is the only technique, used for flaky test detection.\\
\textbf{Hard coding}: Developers change test cases without proper approval and write If-else, in order to make test pass, but it leads to TF, when product changes as stated by one of the participants \textit{"most of the time, if we have determined if test case is flaky [...]  it is usually we wrote a code to a function on a specific property of the hardware, for example, it should work on 640x480 resolution and then new product come that does not have this and they make small changes in hardware and then [tester] hardcoded this test case [if-else]. This is not smart"}. Another participant elaborated that these hardcoded if-else grow over time and make test case so complicated that you forget what test case was actually testing.\\
\textbf{Unstable continuous integration release}: Continuous integration platforms play a major role in TF as mentioned by of the participants \textit{"on the Jenkins side, the biggest problem, they had is flakiness in Jenkins it was that when a new release of Jenkins is used, and this is invisible to user because user does not care what version of Jenkins they are using so for them it is flaky because, they are ignoring the Jenkins version and they are running the test and suddenly nothing is working and they do not know what happened [....]"}\\
\textbf{Web component run-time generation}: One of the case company develops web-based products and they experience flakiness due to web-components as mentioned by one of the participants \textit{"I would say custom web component is also a flaky issue. if you have a test running against a web page, if you write that yourself, you can use Ids but if you use custom web component, most of them are created on the fly when generating the page and those components are very hard to make robust"}\\
\textbf{Complex feature avoidance}: Some practitioners labeled test cases as flaky because it tests a complex feature/requirement. If we can avoid testing this specific feature/requirement, we can avoid test flakiness as mentioned by one of the participants \textit{"we are ignoring X and then we avoid complex feature testing"}. One can argue about trade-offs between risking for less test coverage and reducing test flakiness but this is not a scope of this paper.
\subsection{Quantitative Results}
\label{quatitativeresults}
\begin{figure*}
	\centering
	\includegraphics[width=\textwidth]{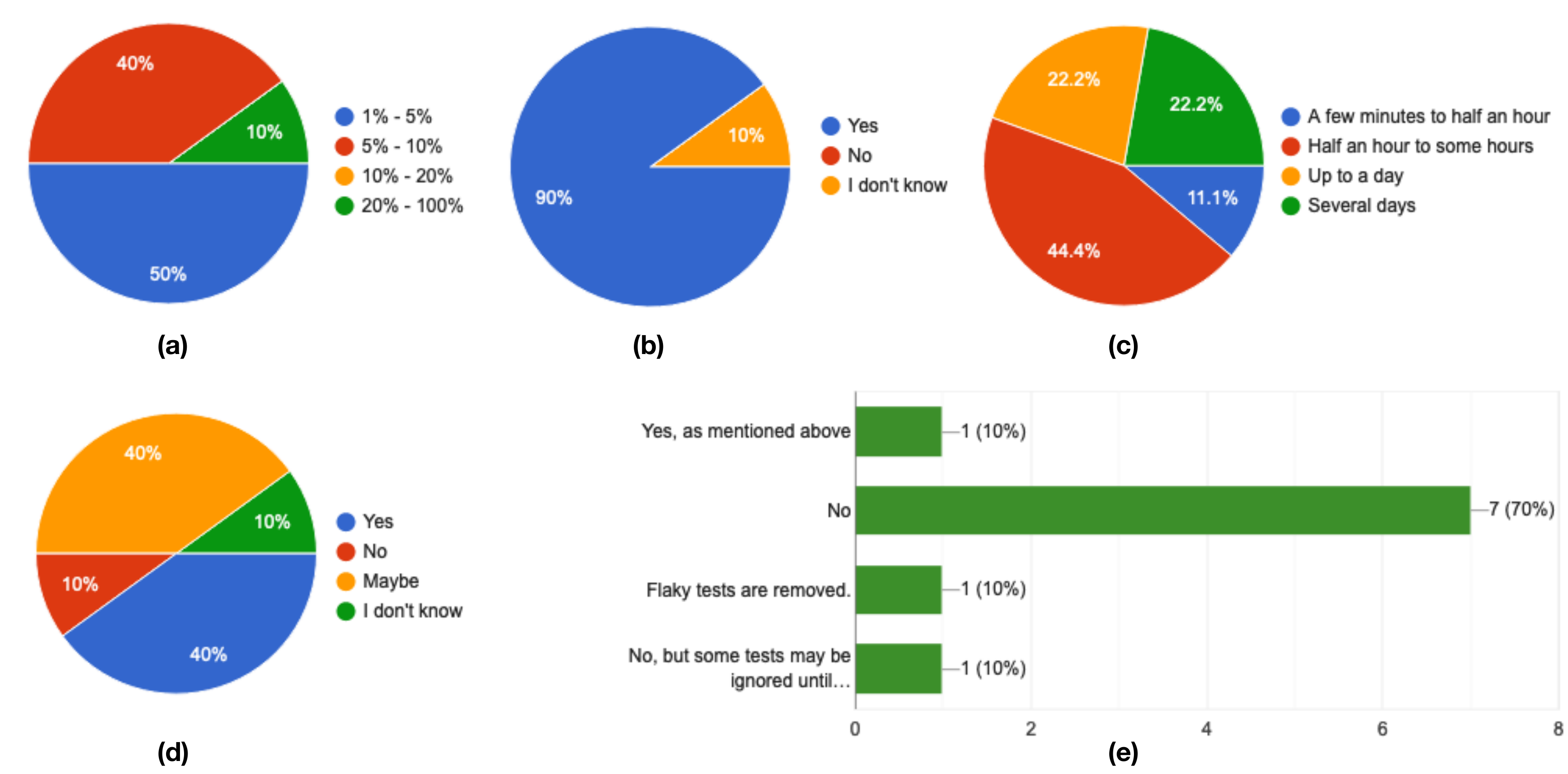}
	\caption{Initial survey results of 10 participants from 4 different companies. (a) What is the rate of test cases that you believe exhibit flaky behavior?. (b) Do you have any review process for writing a test case? (i.e. Does anyone, other than the person who wrote the test case, review the test case). (c) How much time do you spend resolving test case flakiness for each suspected case?. (d) Is this true, in your observation, that most flaky test are flaky from the time they are written. (e) If you notice a test case as flaky, do you use any type of annotation such as "@FlakyTest", "@Repeat", "@Ignore" or "@ReRunThis". }
	\label{fig:quanti}
\end{figure*}
This section presented a few results from the initial survey which was conducted with interested participants at the start of this study. Figure\ref{fig:quanti} represents results of the three questions, mentioned in the survey \footnote{\url{ https://docs.google.com/forms/d/e/1FAIpQLSeOk8YswlfdeEmmRJsVZRCYI_0zwQiRb1RGOinXKVLpkyovew/viewform}}. 5, out of 10, participants shared that their flaky tests are accounted for 1\% -  5\% while other 4 participants have 5\% - 10\% flaky tests (Figure \ref{fig:quanti}(a)) while executing tests. These may appear small percentages but if case companies have many thousands of test cases, then these percentage matters All companies (e.g., 9 out of 10 participants said yes in Figure \ref{fig:quanti}(b)) have test review process but still flaky tests can sneak into the test suite. This represents that their test design process doe not include guidelines to avoid TF. More than 5 participants, out of 10, shared that they spent half an hour to up-to-the day (Figure \ref{fig:quanti}(c)) to resolve test case flakiness. This time does not include the administration hours such as creating bug reports or assigning issues to users. In addition to the different perception about what is TF, we can see in (Figure \ref{fig:quanti}(d)) that participants have answered differently when we asked the question \textit{"Is this true, in your observation, that most flaky tests are flaky from the time they are written?"}. 40\% said 'yes' while the same amount opted for 'maybe'. 7 out of 10 participants shared that they do not use any notations to represents flaky tests (Figure \ref{fig:quanti}(e)). We do not intend to provide all responses to the questions of the survey due to the scope of this paper.
\section{Evaluation}
\label{evaluation}
This section presents two evaluations of the identified factors (1) and their effect on test flakiness (2). We conducted both evaluations separately but on the same day.
\subsection{Evaluation of Identified Factors}
\label{evaluation1}
We conducted this evaluation with four different companies to understand how important these individual factors are for practitioners before evaluating factor's effect. During the first evaluation, the identified factors were ranked by the participants (These participants are different from those participated in data collection). It was the first occasion, they learned about these factors. Each factor was ranked with respect to their importance in terms of high, medium and do not agree. We replaced 'low' with 'do not agree' because a part of the research is to capture developers' perceptions about TF and it is important to know if they do not agree with any of the factors. Each factor was discussed among all participants to avoid any confusion. Figure 3 represents total agreement score, assigned to each factor by practitioners. The agreement score was calculated using the equation \ref{eq:evaluation1}. We have achieved 84\% average agreement rate with respect to identified factors as shown by red line in Figure 3. We have also calculated median of 86\% which complements our overall average. The reason to calculate mean is due to the fact that some researchers do not prefer average on Likert Scale. The lowest agreement rate of 43\% and 50\% are assigned to unstable CI release and web component run-time generation respectively. We expect this ranking because what developers perceive test flakiness in their organization may not be applicable to other developers in other workspace.  
\begin{equation} 
\label{eq:evaluation1} 
S=\frac{(N_H*1)+(N_M*0.5)}{N_H+N_M+N_D}*100
\end{equation}
\newenvironment{conditions}
{\par\vspace{\abovedisplayskip}\noindent\begin{tabular}{>{$}l<{$} @{${}={}$} l}}
	{\end{tabular}\par\vspace{\belowdisplayskip}}
where:
\begin{conditions}
	N    &  Number of Participants \\
	H , M, D & High, Medium or Do Not Agree \\
	1 & Weight assigned to High\\
	0.5 & Weight assigned to Medium\\
	0 & Weight assigned to Do Not Agree
\end{conditions}
We also observed that only one participant ranked 6 factors (e.g., undermining network infrastructure, higher dependency among test cases, complex feature avoidance, hard coding, team motivation and robust test cases) as 'do not agree' and we assumed that the understanding level of the participants might not be on the similar level due to the fact that 'higher dependency among test cases' is a known factor that contributes to test flakiness but still this participant ranked it as 'do not agree'. Since the survey in the workshop was anonymous, we had no means to talk to specific participants to get clarifications. 
\subsection{Evaluation of Factor's Effect on Test Flakiness}
\label{evaluation2}
\begin{figure*}
	\centering
	\includegraphics[width=\textwidth] {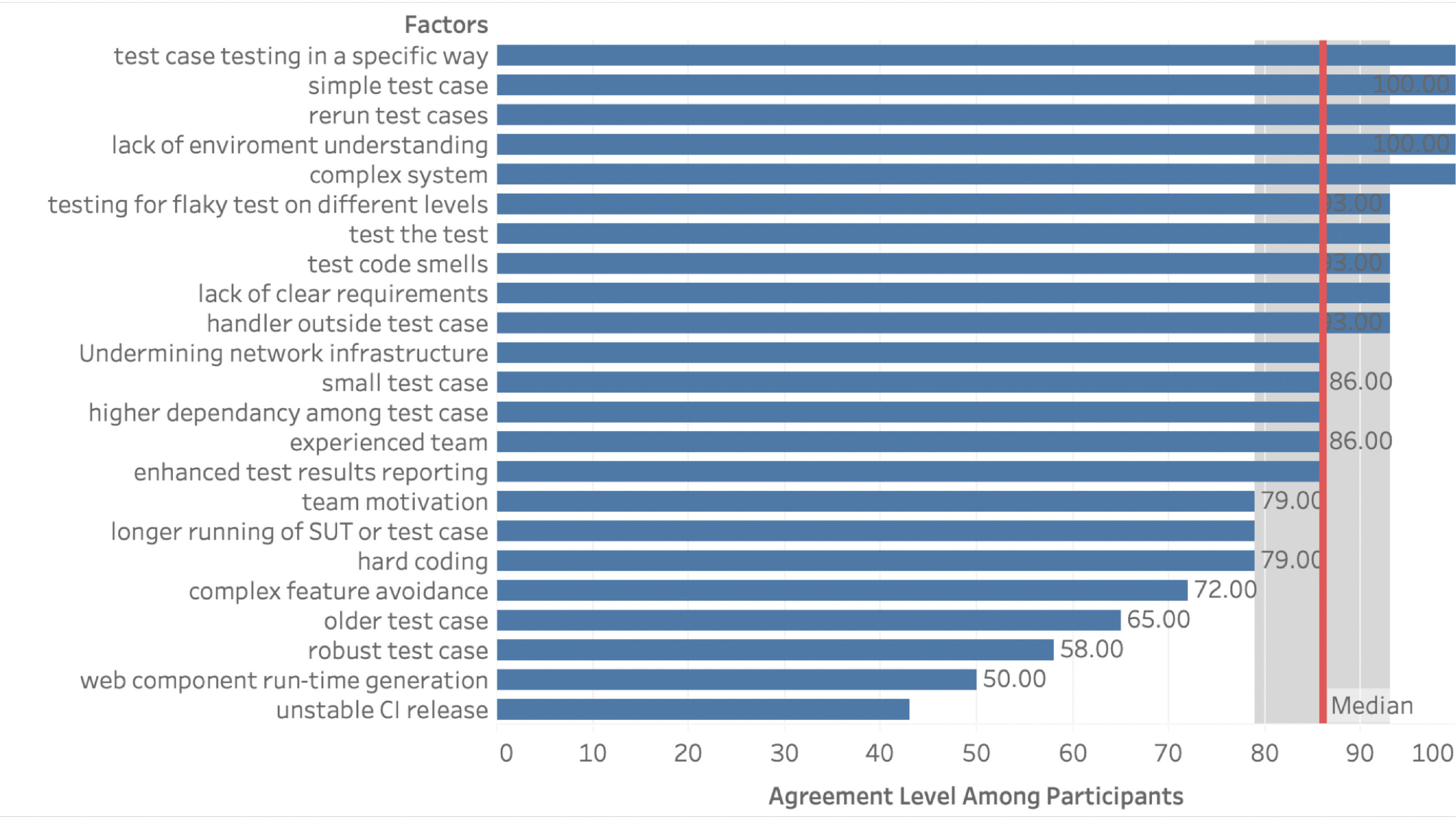}
		\caption{Agreement Level in Percentage of Factors by Participants}
	\label{fig:factors}
\end{figure*}
We assume that survey participants have a good understanding of all factors after the first evaluation. We conducted the second evaluation to rank the effect of identified factors on test flakiness. Figure 4 represents an evaluation of the factor's effect on test flakiness. Participants were asked to rank each effect of a single factor with respect to Likert \footnote{https://www.simplypsychology.org/Likert-scale.html} scale. Since these effects are based on identified factors and participants have already ranked those factors earlier, we expected a similar trend as in Figure 3. Figure 4 represents total agreement score, assigned to each factor's effect by practitioners. The agreement score was calculated using the equation \ref{eq:evaluation2}. We have achieved 86\% average agreement rate with respect to identified factor' effect as shown by red line in Figure 4. We have also calculated median of 89\% which complements our overall average. 
We observed only one case of 'Strongly Disagree' and three cases of 'Disagree'. The effect (e.g., unstable CI release => + TF) was ranked 'Strongly Disagree' by only one participant and this is expected due to the fact that this is a company-specific factor. Other three factors (e.g., testing for flaky test on different level =$>$ - TF, older test case =$>$ + TF, longer running SUT or test case = $>$ * TF) were ranked as 'Disagree' by only one participant. We hold the same opinion that one participant did not understand the factors or their effect.  
\begin{equation} 
\label{eq:evaluation2} 
S=\frac{(N_{SA}*4)+(N_A*3)+(N_N*2)+(N_D*1)}{N_{SA}+N_A+N_N+N_D+N_{SD}}*100
\end{equation}
where:
\begin{conditions}
	N    &  Number of Participants \\
	SA, A, N & Strongly Agree, Agree, Neutral \\
	D, DA & Disagree, Strongly Disagree \\
	4, 3, 2 & Weight assigned to SA, A, N receptively\\
	1, 0 & Weight assigned to D, SD receptively\\
\end{conditions}
\begin{figure*}
	\centering
	\includegraphics[width=\textwidth]{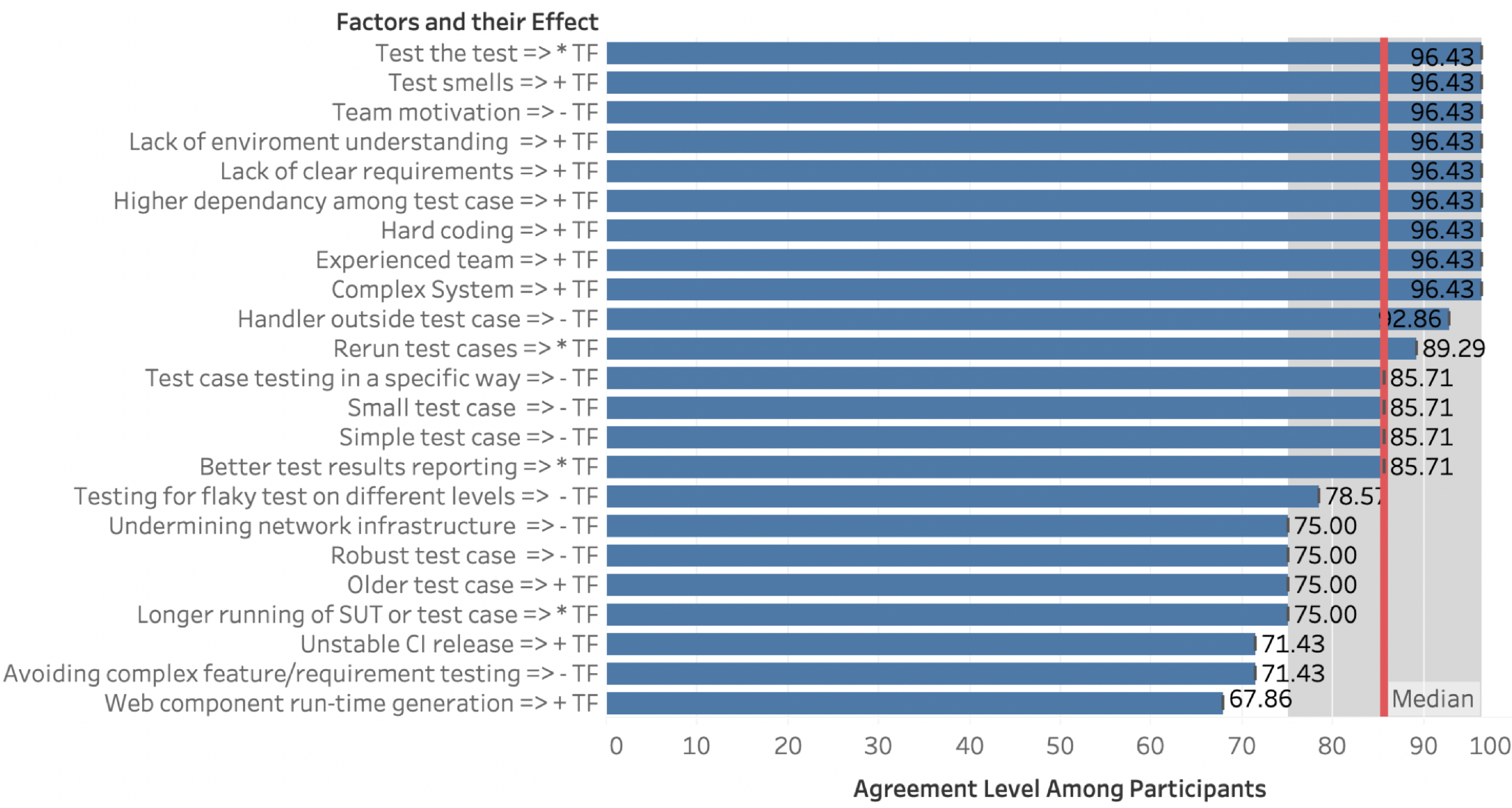}
	\caption{Ranking of the Affect of Identified Factors on Test Flakiness by Participants }
	\label{fig:affect}
\end{figure*}
\section{Discussion}
\label{discussion}
This section will discuss the difference in the practitioner's perception of factors related to test flakiness. We have observed different perceptions from the very beginning when we conducted the online workshop and site visit for data collection. Different participants provided different perception for what flakiness is and whether we should call it test flakiness, source code flakiness or environment flakiness. One of the participants stated \textit{"It is a responsibility of the test to not to be flaky, even if you have flakiness in the SUT or something else because in my definition, you are not changing any code and you are running the test and test is giving you different results [....]I think this is kind of important, I hope it will not take too much time as this can be philosophical discussion about what is flakiness. You can define the flakiness as a point of the observer [....]"}. We have observed that some of the participants relate the word test flakiness with any flaky behavior (e.g., if the system under test is flaky). In their opinion, the test should be able to catch any flaky behavior anywhere in the testing environment. This lead to a very basic question of a test case scope. What is the responsibility of a test case and what should not be expected from the test case? We noticed that practitioners are aware of the scope problem as mentioned by one participant \textit{"someone made a temp board which is not actual hardware. The temp bard will not show exact behavior as hardware does and some tests are really unforgiving towards this because they are not developed for these temp boards. What needs to be done for this is to define the scope of hardware [...] so for them, again it is not flaky behavior if you just look at correct hardware"}.

In addition to what is test flakiness and where it originates from, we observed that no one takes responsibility if test flakiness is detected. Due to the lack of understanding of test flakiness, most practitioners believe in blame-others strategy as mentioned by one of the participants \textit{:Yes, it is usually back and forth between us and developers as they say that this flakiness is due to you and we say that this flakiness us due to you [...] it must be fixed by them"}. It has been observed that upon detection of test flakiness, it is deleted, skipped or ignored from the test suite and on a very different occasion, the team actually hunts for the root cause of test flakiness.

We have also observed differences in perceptions of factors among practitioners when we conducted the workshop to validate our findings.  The factor 'robust test case' was discussed for a longer time among participants where one participant did not agree that this factor affects test case flakiness while two participants mentioned it as high and three mentioned it as medium. We observed another extreme case with the factor '*unstable CI release' where three participants mentioned it as 'high' and four mentioned it as 'do not agree'. Although this factor is specific to one company, but upon discussion, participants from another company also agreed to this factor. Another factor that received equal ranking towards 'high', 'medium' or 'do not agree' is web component run-time generation. This was specific to one company and participants from this company strongly consider this factor to be a major cause of test flakiness while most of the participants did not agree with it. This validates our definition of perception (e.g., see Section \ref{background}) where we stated that their perception, about test case flakiness, is more likely to influenced by what they have heard, observed and experiences in the workplace.

We have observed that all the identified factors have been a part of the discussion in the literature in different contexts (see Table IV). The identified factors such as 'simple test case', 'small test case' and 'test case testing in a specific way' have been mentioned as 'simplicity' and 'single responsibility' in \cite{bowes_how_2017} as part of the properties of a good test case. The identified factor 'handler outside test case' is called as 'simplicity of fixtures' in \cite{bowes_how_2017}\cite{rompaey_characterizing_2006}. A robust test case which was discussed longer during the evaluation of results (see Section \ref{evaluation}) is mentioned with the same name in \cite{deursen_refactoring_2001}. More mapping can be seen in Table IV. We concluded that what practitioners perceive as factors affecting test flakiness are actually embedded in the following points and can be avoided if they focus on good quality test cases:
\begin{enumerate}
	\item How good their test cases are?
	\item What are the design and review guidelines for writing and evaluating test cases?
	\item Do their design and review guidelines address test flakiness prevention?
	\item How to practitioners perceive the quality of software and does flakiness is considered a threat to the product quality?
	\item Flaky tests are just a bad test case.
\end{enumerate}
\section{Validity Threats}
\label{validtythreasts}
We analyzed validity threats in this section that may affect the outcome of this study \cite{runeson_case_2012}.
\subsection{Internal Validity}
An internal validity threat can be that participants did not understand our coding and its representation correctly which we also observed as one participant ranked important and known factors as do not agree as described in Section \ref{quatitativeresults}. We tried to reduce this to conduct a workshop in person at our University so we can explain these factors and their effect. In addition to this, we dedicated some time for questions, if any, related to these factors and their effect.
\subsection{Construct Validity}
We have eliminated constructive validity threat completely by conducting both workshops (e.g., data collection and data validation) with different participants. We eliminated the participant's bias (e.g., pleasant answer) by conducting two workshops. We have also eliminated the researcher's bias by involving all 3 researchers during the design of the workshop and questions.
\subsection{External Validity}
External validity refers to what extent it is possible to generalize the findings as well as to what extent the findings are of interest to other practitioners other than the investigated case \cite{runeson_case_2012}. We tried to eliminate external validity threat by selecting four different companies that work in different domains. Since we have coded some company-specific factors (e.g., see Section \ref{companyspecific}), we cannot eliminate external validity completely.

\section{Conclusion}
\label{conclusion}
The presence of flaky tests in the test suite raises concerns over product quality. It affects the confidence level whether my products is up to the better quality or not. It is very important that practitioners and researchers focus on techniques/tools/methods/guidelines/approaches that prevent TF rather than detecting it after test suite execution similar to any disease where we take precautionary measurements before it spreads. We have conducted an online survey, an online workshop and the site visit in four different companies. We have identified 23 factors that increase, decrease or affect the ability to detect test flakiness. We concluded that some factors can be combined together to have an adverse affect on TF. We evaluated our results with different participants and presented our findings. We have also concluded that practitioners have different perception about what is test flakiness and where does it lie in the continuous integration pipeline. What also captured the practitioner's perception about the factors that have an effect on test flakiness. We concluded that there is a need to define what is test flakiness and how to avoid it.


\section*{Acknowledgment}

The authors would like to thank  the participants in our reported multiple case study for their availability to help us in data collection, data validation and clarify questions about the data as well as the engaging and insightful discussions. This work was supported by the Linköping Software Center \footnote{www.software-center.se} Project 30 on Aspects of Automated Testing.



%

\bibliographystyle{unsrt}
\bibliography{references.bib}

\end{document}